\documentclass[a4paper,twocolumn,11pt,
unpublished
]{quantumarticle}

\pdfoutput=1

\usepackage[utf8]{inputenc}
\usepackage[english]{babel}
\usepackage[T1]{fontenc}
\usepackage{amsmath}
\usepackage{amssymb}

\usepackage{hyperref}

\usepackage[square,numbers,sort&compress]{natbib}
\usepackage{graphicx}
\newcommand{\beq}{\begin{equation}}
\newcommand{\eeq}{\end{equation}}
\newcommand{\eq}[1]{\eqref{#1}}

\newcommand{\bra}[1]{\langle #1|}
\newcommand{\ket}[1]{|#1\rangle}
\newcommand{\vac}{|\mathrm{vac}\rangle}

\newcommand{\myrule}{\rule[-2pt]{0pt}{2pt}}

\newcommand{\create}[1]{a_{#1}^\dagger}
\newcommand{\destroy}[1]{a_{#1}}
\newcommand{\createN}[2]{a_{#1}^{\dagger #2}}
\begin{document}

\title{Fock Space Perspective on Optimal Heralding Schemes}

\author{Gubarev F.V.}
\affiliation{DATADVANCE LLC, Russia 117246, Nauchny pr.~17, 15~fl., Moscow}
\affiliation{Quantum technology centre, Faculty of Physics, Lomonosov MSU, Moscow, Russian Federation}

\maketitle
\begin{abstract}
We address the design of optimal heralding schemes in linear optical quantum computing context.
The problem admits unit fidelity formulation in relevant Fock space,
thanks to linear optic feasibility criteria of Ref.~\cite{PhysRevA.100.022301}.
Corresponding solution methodology is presented.
As an application we inspect optimality of a few known schemes of Bell states generation.
It is shown that in case of six modes and two ancillary photons success rate 2/27 is suboptimal.
\end{abstract}

\section{Introduction}
\label{section::intro}

Efficient heralding of desired multi-photon quantum states is one of
the corner stones of linear optical schemes based on KLM protocol~\cite{Knill2001ASF}.
It could be summarized as follows (see, e.g.,
Refs.~\cite{Knill2002QuantumGU,MyersLaflamme2006,RevModPhys.79.135,OBrien1567,TAN2019100030,Flamini_2018}
for detailed exposition).
Predefined number-basis state
$\ket{\psi_{in}} = \ket{\vec{n}} = \ket{n_1 \dots n_N}$
of $n = \sum_i n_i$ photons in $N$ modes is subjected to passive linear optical transformation $U(S)$,
parameterized by $N \times N $ unitary scattering matrix $S$ pertinent to underlying interferometer.
$U$ acts in Fock space of dimensionality
{\small $N_{st} = \left(\begin{array}{c} N + n - 1 \\ n \end{array}\right)$}
and is of corresponding size.
Then partial measurement $\bra{\vec{m}} U \ket{\vec{n}}$
of $m = \sum_j^M m_j$ photons in last $M$ modes is performed.
The problem is, for given target state
$\ket{\psi_{out}} = \sum \alpha_{\vec{k}} \ket{\vec{k}}$ of $n-m$ photons in $N-M$ modes,
to find unitary $S$, for which successful measurement heralds exactly the desired wave function,
$\ket{\psi_{out}} \propto \bra{\vec{m}} U \ket{\vec{n}}$ with maximal success
probability $P = |\bra{\vec{m}} U \ket{\vec{n}}|^2$.

It is apparent that posed optimization task hardly admits analytical solution even for
moderately sized geometries, to say nothing of optimality proofs.
Numerical insights seem necessary and traditional treatment proceeds via particular parameterization of $S$
(see, e.g.,
Refs.~\cite{Clements:16,Reck1994ExperimentalRO,Guise2018SimpleFO,Diele1998TheCT,Jarlskog2005ARP}).
Since complete matrix elements are given by permanents of various matrices \cite{Scheel2004PermanentsIL}
formed from $S$, one could try to maximize the success probability $P$
subject to unit fidelity constraint
$ F = | \sum_{\vec{k}} \alpha_{\vec{k}} \bra{\vec{k},\vec{m}} U \ket{\vec{n}} |^2 / P = 1$.
However, in this form the problem appears numerically awkward mostly because of
required maximal overlap,
though evaluation of permanents and corresponding strong non-linearities are important as well.
Unit fidelity is essential, however, respective function never exceeds one.
Therefore, usual assumption (see, for instance, \cite{NoceWrig06}) of constraint qualifications does not hold,
which hinders application of conventional constrained optimization algorithms.
An alternative might be to explicitly impose
$\bra{\vec{k},\vec{m}} U \ket{\vec{n}} = \sqrt{P} \alpha_{\vec{k}}, \forall \vec{k}$,
but, to the best of our knowledge, this approach had never been investigated.

Conventional solution~\cite{Uskov2009MaximalSP,PhysRevA.96.043861,Gubarev2020ImprovedHS,fldzhyan2021fivemode}
is to avoid constrained formulations altogether and resort to a kind of ``merit function'' approach,
which combines into one performance measure both the success probability and the fidelity of outcome.
Popular choice \cite{PhysRevA.96.043861} is to maximize $P F^p$ with some (large) positive power $p$,
which penalizes small fidelity values.
Apart from non-universality of $p$-parameter (its appropriate value is strongly case-to-case dependent),
evident drawback is that unit fidelity is not assured even for successfully
finished runs, to say nothing of vast complication of objective landscape.
As a consequence, it might become difficult to extract reliable information
from outcomes of performed numerical studies.

For us it is essential that passage
to larger design space might often mitigate
inherent non-linearities and make numerical optimization more robust.
In present context it is thus tempting to consider ``large'' unitary $U$ as a prime design variable.
Indeed, unitary transformations are straightforward in entire Fock space,
permanents and related complexities are avoided.
Furthermore, requirement of unit fidelity might be accounted for exactly (Section~\ref{section::heralding-residual}).
On the other hand, only tiny part of all such unitaries are realizable with linear optics,
one needs to impose additional constraints to ensure optical feasibility.
Hopefully, Ref.~\cite{PhysRevA.100.022301} proposed explicit
criterion to decide whether a given unitary $U$ is realizable with linear optics
(we elaborate on this in Section~\ref{section::linear-optics}).
Therefore, it becomes possible to address posed optimization problem from different perspective.
We report the details of corresponding implementation in Section~\ref{section::numerics}.
Finally, in Section~\ref{section::applications} we present a few application examples,
which are confined to relatively small cases of 5 and 6 modes,
but are still sufficiently representative to illustrate
both strong and weak aspects of proposed methodology.
In particular, we confirm apparent optimality of recently proposed~\cite{fldzhyan2021fivemode}
5-port scheme of Bell states generation.
In case of six modes we demonstrate suboptimality of well-known scheme~\cite{Carolan711},
which delivers $2/27$ success probability,
improved circuit exhibits $\sim 5\%$ better performance.
Although the result is of little practical value,
it seems to best illustrate the potential power of proposed methodology.

\section{Fidelity Constraints}
\label{section::heralding-residual}

For fixed measurement pattern $[\vec{m}]$ vectors
$\ket{\mu_\alpha} = \ket{ \vec{k}, \vec{m} }$, $\forall \vec{k}$
provide a basis in space of all heralded states.
Desired output $\ket{\psi_{out}}$ thus obeys
$\ket{\psi_{out}} \otimes \ket{\vec{m}} = a_\alpha \ket{\mu_\alpha}$, $|a|^2 = 1$,
where (and throughout the paper) sum of repeated indices is implied.
If the initial state $\ket{\vec{n}}$ is numbered as first basis vector,
unit fidelity requires $\{\mu_\alpha\}$ subset of elements of
first $U$ column to be aligned with given $a_\alpha$
\beq
\label{heralding-residual}
(1 - a a^\dagger)_{\alpha\beta} \, U_{\mu_\beta, 1} ~=~ 0\,.
\eeq
Corresponding proportionality coefficient
\beq
\label{heralding-amplitude}
U_{\mu_\alpha,1} ~=~ z \, a_\alpha
\eeq
is the success amplitude of target state heralding.

For unit fidelity matrices $U$ equation \eq{heralding-residual} is invariant
under $U \to U g$ with $g$ having zeros in non-diagonal elements of first row/column.
It also remains intact with respect to $U \to U e^{i H}$ with
\beq
\label{nullspace-H-matrix}
H ~=~ \left[\begin{array}{cc} 0 & (\xi \, h ~+~ P_u q)^\dagger \\ \xi \, h ~+~ P_u q & 0 \end{array}\right]\,,
\eeq
where $\xi$ and $q$ are arbitrary complex number and vector, respectively, $h$ is a particular solution
(with somewhat artificial but convenient phase)
\beq
\label{nullspace-h-vector}
h_n = -i \, \frac{z}{|z|} \,a_\alpha \, U^*_{\mu_\alpha, n }\,,\quad n=2,\dots,N_{st}
\eeq
of inhomogeneous equations $U_{\mu_\alpha, n} h_n \propto a_\alpha$
and $P_u$ is an orthogonal projector onto the complement of
$V = \mathrm{Span}\{e^{(\alpha)}\}$, $e^{(\alpha)}_n = U^*_{\mu_\alpha,n}$.
Note that $P_u q$ provides general solution in homogeneous case, $U_{\mu_\alpha, n} [P_u q]_n = 0$.
In fact, these transformations exhaust the symmetries of \eq{heralding-residual},
moreover, unit fidelity solutions are of the form
\beq
\label{nullspace-U-transform}
U ~=~ U^{(0)} \, e^{iH} \, g\,,\quad
g = \left[\begin{array}{cc} e^{i\varphi} & 0 \\ 0 & \Omega \end{array}\right]\,,
\eeq
provided that reference matrix $U^{(0)}$ fulfills \eq{heralding-residual}.
As far as Eq.~\eq{heralding-amplitude} is concerned,
representation \eq{nullspace-U-transform} retains its form, but changes heralding amplitude for $H \ne 0$.
Infinitesimally we have $ z \to z \, (1 + \xi \, (1 - |z|^2)|z|)$, but at higher orders $P_u q$ term also contributes.

Equations \eq{nullspace-H-matrix}-\eq{nullspace-U-transform} provide a basis for developed
approach and allow to maintain fidelity constraints \eq{heralding-residual} exactly.
Within the optimization context generated search direction (Hermitian matrix)
is to be orthogonally projected onto the tangent space of \eq{heralding-residual}
and this affects only non-diagonal elements of first row/column.
Respective projector $\Pi$ retains only the components orthogonal to $V = \mathrm{Span}\{e^{(\alpha)}\}$
or proportional to $h \in V$
\beq
\label{null-space-projector}
\Pi \left[\begin{array}{cc} \varphi & t^\dagger \\ t & \omega \end{array}\right] =
\left[\begin{array}{cc} \varphi & (\xi h + P_u q)^\dagger \\ \xi h + P_u q & \omega \end{array}\right]
\eeq
Then Eq.~\eq{nullspace-U-transform} performs non-infinitesimal change of current unitary $U$,
which still exhibits unit fidelity with target state.

\section{Linear Optics Feasibility}
\label{section::linear-optics}

Next we discuss realizability criterion of unitary $U$ by linear optics.
Already for dimensional reasons one expects that it should be equivalent
to large set of constraints, which seem to were unknown until the work~\cite{PhysRevA.100.022301}.
Hopefully, Ref.~\cite{PhysRevA.100.022301} addresses the issue in full and provides suitable formulation,
which we first overview (albeit in a bit different language) and then indicate how to incorporate
it in present context.

Scattering matrix $S$ might be considered as an element of SU(N), $S = e^{i s_a t^a}$,
where $t^a$ are conventional generators.
Respective Fock space transformation is then $U = e^{i s_a T^a}$ with
$T^a = t^a_{ij} \, \create{i} \destroy{j}$
(bilinear combinations of creation/annihilation operators).
Crucial observation of Ref.~\cite{PhysRevA.100.022301} is that the space $W = \mathrm{Span}\{T^a\}$ is $U$-adjoint invariant
\beq
\label{optic-feasibility-original}
U \, T^a \, U^\dagger ~ \in ~ W\,, \quad \forall a
\eeq
and that this condition is sufficient for $U$ to be realizable with linear optics.
We'll utilize slightly different form of \eq{optic-feasibility-original}
using normalized sparse traceless real matrices
\beq
\label{gamma-def}
\gamma^{ij} = ( \,\create{i}\destroy{j} - \frac{n}{N} \delta^{ij} \, ) \,/\, \sqrt{2K}
\eeq
with $K = n (N + n) N_{st}  / [2 N (N + 1) ]$, which obey
\beq
\label{gamma-trace}
\mathrm{Tr}[ \gamma^{ij} \gamma^{nm} ] ~=~ \delta^{im} \delta^{jn} - \frac{1}{N} \delta^{ij} \delta^{nm}\,,
\eeq
\beq
\label{gamma-commutator}
\sqrt{2K} \,\, [\gamma^{ij},\gamma^{nm}] = \delta_{jn} \gamma^{im} - \delta_{im} \gamma^{nj}\,,
\eeq
together with their respective ``rotated'' dense counterparts $\bar{\gamma}^{ij} = U^\dagger \, \gamma^{ij} U$.
Optical feasibility condition \eq{optic-feasibility-original} could then be represented as
\beq
\label{optic-feasibility-projectors}
(1 - \Gamma) \, \bar{\Gamma} = (1 - \bar{\Gamma}) \, \Gamma = 0\,,
\eeq
where $\Gamma$ and $\bar{\Gamma}$ are orthogonal projectors, constructed from ``bare'' and ``rotated'' $\gamma$-matrices
\beq
\Gamma_{\mu\nu, \lambda\rho} ~=~ \gamma^{ij}_{\mu\nu} \, \gamma^{ij}_{\lambda\rho}\,,\quad
\bar{\Gamma}_{\mu\nu, \lambda\rho} ~=~ \bar{\gamma}^{ij}_{\mu\nu} \, [\bar{\gamma}^{ij}_{\lambda\rho}]^*
\eeq
and viewed as matrices of size $N^2_{st} \times N^2_{st}$.
Thus realizability of given $U$ by linear optics, Eq.~\eq{optic-feasibility-projectors},
requires that two bases
$\gamma^{ij}$ and $\bar{\gamma}^{ij}$ define the same linear subspace and hence are unitary related
\beq
\label{unitary-equivalent}
\bar{\gamma}^{nm} \, G^{(nm),(ij)} ~=~ \gamma^{ij}\,.
\eeq
Here $G$ is $N^2 \times N^2$ unitary matrix with respect to indicated twin indices,
explicit form of which follows from \eq{gamma-trace}
\beq
G^{(nm),(ij)} = \mathrm{Tr}[ \gamma^{ij} \, \bar{\gamma}^{mn} ] + \frac{\delta^{nm} \delta^{ij}}{N}\,.
\eeq
Confronting \eq{gamma-def}, \eq{unitary-equivalent} with conventional transformation rules of
creation/annihilation operators, one obtains
\beq
\label{unitary-equivalent-scattering}
G^{(nm),(ij)} = S_{ni} S_{mj}^*\,,
\eeq
where $S$ is $U$-realizing scattering matrix of underlying interferometer,
which could be extracted (up to the global phase) from
\eq{unitary-equivalent}-\eq{unitary-equivalent-scattering}.

In practical implementations, it seems better to consider optical feasibility conditions
as zero-normalized equality constraints
\beq
\label{optic-feasibility-criterion}
R^a ~=~ (1 - \Gamma) \, U T^a U^\dagger ~=~ 0\,,
\eeq
to which we refer as optical residuals; the term ``optical residual'' $R$
is used for respective squared $L_2$-norm, which up to irrelevant constant reads
\beq
\label{optic-residual}
R = 1 - \frac{1}{N^2 - 1} \, |\, \mathrm{Tr}[ \gamma^{ij} \, \bar{\gamma}^{nm} ] \,|^2 \propto |R^a|^2\,.
\eeq
If there would be no fidelity constraints and no need to maximize success probability
an efficient approach to solve \eq{optic-feasibility-criterion} is to consider
iterative Newton-Raphson scheme $U \to U e^{i\tau X}$, in which Hermitian
$X$ approximately solves $J^a \, X = - R^a$, while \eq{optic-residual}
is used in line searches to establish appropriate $\tau$ and assure global convergence.
Here $J^a$ denotes the Jacobian of \eq{optic-feasibility-criterion}.
As discussed above, unit fidelity requires projection onto the tangent space of fidelity constraints.
Therefore, Newton-like ``normal'' step $H_N$ is to be determined from 
linear system $J^a \, \Pi \, X = - R^a$, normal form of which
\begin{gather}
\label{HN-normal}
\Pi \, (J^{a\,\dagger} J^a) \, \Pi \,\, X = -\Pi \, J^{a\,\dagger} R^a\,, \\
H_N = \Pi X \nonumber
\end{gather}
we solve with unpreconditioned conjugate gradients (details are provided in Section~\ref{section::numerics}).
On the other hand, maximization of success rate requires to identify tangent space of optical residuals as well.
In fact, this could be done easily thanks to transparent geometrical meaning of Eq.~\eq{optic-feasibility-projectors}:
its symmetries correspond to infinitesimal transformations $\delta\bar\gamma \propto [\bar\gamma,X]$,
which leave the space $\mathrm{Span}\{\bar\gamma\}$ invariant.
From \eq{gamma-commutator} we conclude that tangents to optical residuals are of the form
\beq
\label{HT-generic}
H_T = \alpha_{ij} \bar{\gamma}^{ij}\,, \quad \alpha^*_{ij} = \alpha_{ji}\,,
\eeq
with otherwise arbitrary complex $\alpha_{ij}$.
However, Eq.~\eq{HT-generic} is not in general consistent with unit fidelity conditions \eq{heralding-residual},
which require $H_T$ to belong to respective tangent space as well.
Therefore, additional linear constraints
\beq
\label{HT-constraints}
(1 - \Pi)\, H_T = 0
\eeq
are to be imposed on admissible $\alpha_{ij}$ coefficients (cf. Eq.~\eq{null-space-projector}).

Note that the number of degrees of freedom $N_{DoF}$ remaining upon the imposition of \eq{HT-constraints},
is the dimensionality of feasible domain.
It might well happen that \eq{HT-constraints} admits only trivial solution $\alpha_{ij} = 0$,
in which case problem is likely to possess isolated feasible points only.
$N_{DoF}$ depends on the considered setup, moreover, one cannot exclude its changes with varying $U$.
However, experience revealed that for generic $U$ later alternative is highly improbable,
feasible domain dimensionality is well defined unless unitary transformation is chosen exceptional.
Remaining $N_{DoF}$ variables are to be exploited to maximize the success rate,
respective Newton-like tangent step is determined essentially by maximal overlap of $H_T$ and
the only non-zero mode \eq{nullspace-h-vector} of success probability
(we drop corresponding straightforward but cumbersome expressions).

\section{Implementation Details}
\label{section::numerics}
Identified normal/tangent spaces allow to treat the problem within SQP-like framework~\cite{NoceWrig06}.
Note that application of available general purpose optimization algorithms is problematic
because of foreseen large dimensionality.
Hopefully, with only equality constraints and analytic knowledge of normal/tangent directions
respective efficient optimization technique becomes rather simple.
Given an initial approximation $U$, which satisfies
\eq{heralding-amplitude}, implemented method might be summarized as follows:
\begin{itemize}
\item Determine normal $H_N$ and tangent $H_T$ Newton-like steps;
\item Select an appropriate positive parameter $\eta$ such that $X = H_N + H_T$ is a descent
  direction for merit function $R - \eta \, |z|^2$, which is to be line searched
  along $X$ via \eq{nullspace-U-transform};
\item Iterate the above until convergence with respect to both optical residual \eq{optic-residual}
  and magnitude of tangent step $H_T$.
\end{itemize}
Note that the above sequence constitutes local optimization algorithm, which might converge
therefore to only locally optimal solution or even stop at infeasible stationary points of optical residual.
This might be cured as usual via repeated restarts with random initial point.

Let us comment on a few important technical details.
First, it seems that explicit large dense matrices are unavoidable, at very least unitary $U$
and ``rotated'' $\bar\gamma$-matrices are to be kept.
This might severely restrict applicability range of proposed methodology: required memory scales
as $N^2_{st} N^2$, with double precision and nowadays ordinary $64$ Gb personal
computers we obtain $N_{st} N \lesssim 10^4$ with uncertain factor on right hand side of order a few.
Then it is apparent that approach is applicable only to relatively
small problems with $\lesssim 10$ number of modes/photons.
Nevertheless it still fits the range of current experimental capabilities.
Besides, even with these limitations our approach is of particular significance since
it allows to check (or at least provide independent evidences of) optimality of purported designs.

The most technically involved stage of the above procedure is the determination of normal step \eq{HN-normal}.
It turns out that linear operator $J^{a\,\dagger} J^a$ consists of two parts:
the first one includes only sparse matrices and is computationally inexpensive
\beq
\frac{N^2 - 1}{N_{st}} X - \gamma^{ij} \, X \, \gamma^{ji}\,,
\eeq
while the second commutator term
\beq
\label{J-commutator}
-\frac{1}{2} \, ( \, \mathrm{Tr}( \, [\bar\gamma^{nm}, \gamma^{ij}] \, X \, ) \, )^* \,
\cdot\, [\bar\gamma^{nm}, \gamma^{ij}]
\eeq
refers explicitly to large dense varying at run time matrices.
To reduce computational overhead we use only a subset of $\bar\gamma^{ij}$ with $i \le j$
and (if we're not too short in memory) precalculate commutators at start of each iteration.
Solution to \eq{HN-normal} is conducted with conjugate gradients using adaptive precision,
dictated by currently achieved optical residual.
With this strategy normal step construction becomes relatively expensive only
close to optical feasibility thus suggesting further performance improvement.
Indeed, our current implementation does not use preconditioning,
however, at small $R$ it is possible to (approximately) apply unitary equivalence \eq{unitary-equivalent}
and greatly simplify the expression \eq{J-commutator}.
It seems that this way an efficient preconditioner might be obtained, but this is still postponed for future work.

Determination of tangent direction $H_T$ reduces to low-dimensional linear algebra
and (semi-)analytic maximization of success rate as a function of single argument
(overlap of $H_T$ with mode \eq{nullspace-h-vector}).
In update rule \eq{nullspace-U-transform} $\Omega$-block is evaluated via
Cayley transform~\cite{Diele1998TheCT,WenZaiwenYinWotao2010,Gubarev2020ImprovedHS},
which requires single inversion of large dense matrix and hence might become prohibitive.
This is addressed via low-rank approximation to $\omega$ with Lanczos algorithm
and adaptively selected number of Lanczos vectors.
Resulting implementation of \eq{nullspace-U-transform} is
much less time consuming than normal step construction and yet sufficiently accurate
for not to spoil algorithm convergence.

\section{Applications}
\label{section::applications}

\subsection{Separable Photons}
\label{section::applications::separable}

We first illustrate developed methodology in simple toy-level test case of ``heralding''
separable photons, where all the results are explicitly known.
Specifically, we consider 4-port optical device, which is injected with 3 single photons,
on output one ancillary photon is measured and the purpose is to herald
one-particle states in second and third ports with maximal probability.
Hence, the problem is defined by
$\ket{\psi_{in}} = \ket{1110}$, $\ket{\psi_{out}} = \ket{011}$, $\ket{\vec{m}} = \ket{1}$.
Optimal success rate is obviously one, relevant interferometer is described
by an appropriate permutation matrix (degenerate multiple optimal solutions).

\begin{figure}[t]
\centerline{\includegraphics[width=0.5\textwidth]{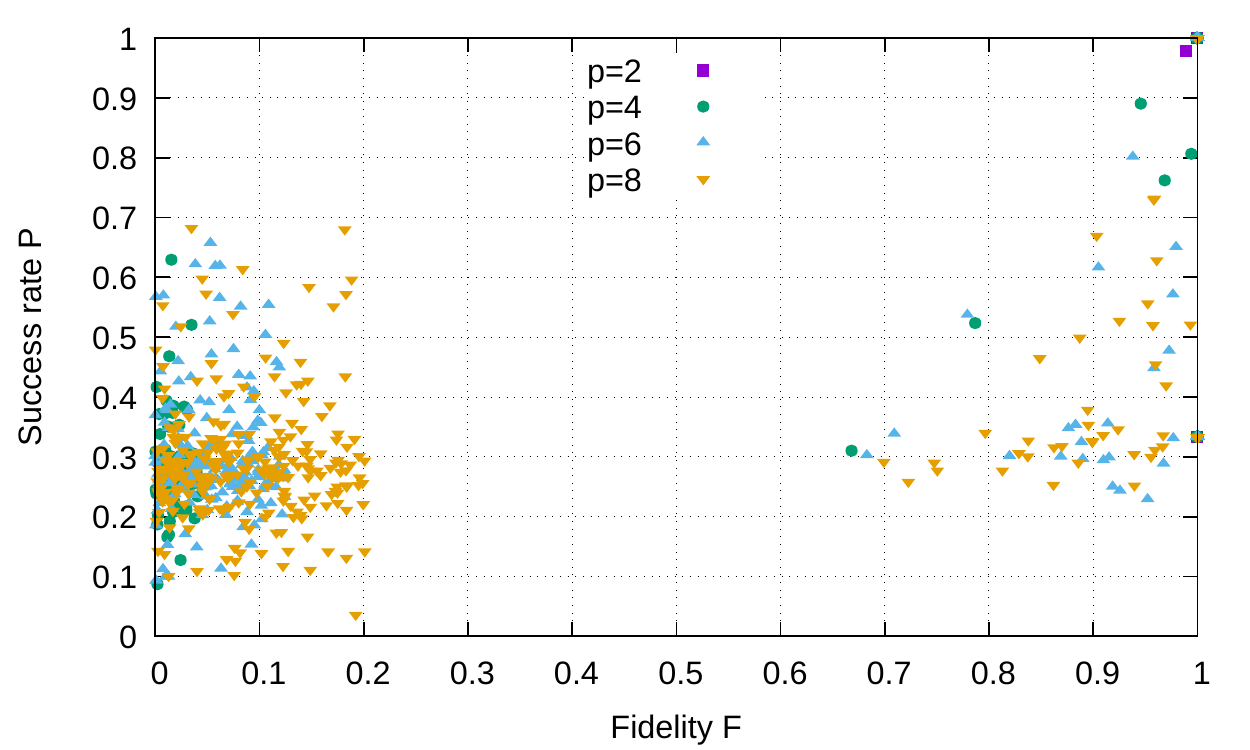}}
\caption{Heralding of separable photons with 4-ports:
fidelity and success rate pairs $(F,P)$, optimized via $\max P F^p$ with indicated
$p$-parameter values. Each set of points corresponds to 400 independent runs conducted
as in Ref.~\cite{Gubarev2020ImprovedHS}.}
\label{fig::cayley-runs}
\end{figure}

Despite of its triviality, example is numerically subtle,
well illustrates the issues discussed in Introduction
and provides an adequate performance estimate of proposed methodology.
In particular, already in this case difficulties of traditional
``merit function''-like treatment appear in full.
If one performs a series of independent optimizations of cumulative performance measure $P \, F^p$,
expected result is obtained in no more than $\sim 75\%$ of cases, success rate
strongly depends on selected parameter $p$ (experiments were conducted along the lines
of Ref.~\cite{Gubarev2020ImprovedHS}, to which we refer for further details).
For instance, Fig.~\ref{fig::cayley-runs}
illustrates resulting distributions of $(F,P)$ pairs for different $p$,
each obtained with 400 independent runs, all of which finished successfully with remaining
gradient norm  $\lesssim 10^{-7}$.
Although the multiplicity of each point is not shown, it is evident that
larger values of $p$ make optimization process less reliable.
Most robust behavior is observed at $p=2$, which always converged to $F=1$,
but with two different probabilities,
either $P = 1$ (70\% of cases) or $P=1/3$ (remaining 30\%).
At larger $p$ the problem gradually losses stability, optimized pairs $(F,P)$ spread more in available ranges.
Hence there is an evident issue with appropriate $p$-value selection, for which no guidance is available
even in this simple case.

\begin{figure}[t]
\centerline{\includegraphics[width=0.5\textwidth]{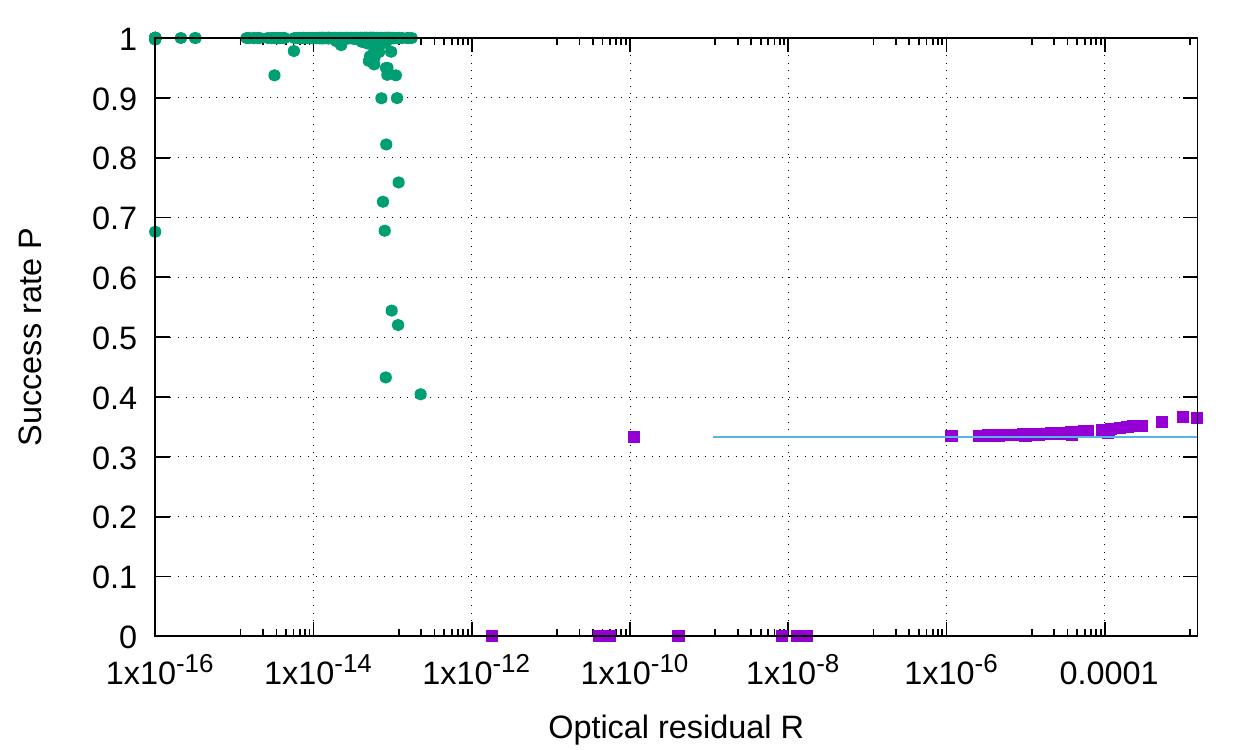}}
\caption{Attained optical residual $R$ and success rate $P$ values
in 400 independent optimization runs for the same problem as on Fig.~\ref{fig::cayley-runs},
horizontal line corresponds to $P=1/3$.}
\label{fig::1::escartin-runs}
\end{figure}

We apply our method to the same problem and performed equal number of independent runs,
convergence tolerances were $\epsilon_R = 10^{-12}$ for admissible optical residual and
$\epsilon_T = 10^{-6}$ for the magnitude of tangent step.
Results are summarized in Fig.~\ref{fig::1::escartin-runs},
which shows all obtained pairs $(R,P)$.
It is apparent that algorithm converges to $(0,1)$ corner in majority of cases,
although there is notable amount of runs, which finished in infeasible position $R > \epsilon_R$.
We attribute the latter to the existence of $R$-stationary point near $P = 1/3$ known
from unconstrained study above.
On the other hand, a few purple dots at $P=0$ correspond to rare cases of inadequately selected
normal step precision, which resulted in non-descent direction and termination of the search.
As for the minor group of green dots with $R < \epsilon_R$ and far from unit probability,
presumably they appear because of slightly inadequate choice of $\epsilon_T$:
optimization seems finished prematurely in feasible domain upon small tangent step attempt.
Of course, this appeal to refinement of utilized termination criteria, but we postpone this task for future work.

As far as specific numbers are concerned, feasible solutions were obtained in 304 cases out of 400,
while near unit ($\ge 0.999$) probability resulted from 282 runs.
We conclude therefore that in the considered toy model proposed method performs similarly
to ``merit function'' approach.
Apparent spread of obtained optimal probabilities might be cured with better convergence criteria.

\subsection{Bell States Generation}
\label{section::applications::bell}

It is known~\cite{PhysRevA.96.043861} that with linear optics
generation of Bell states requires at least four photons,
two respective widely appreciated schemes use either 6 or 8 modes
and have success probabilities
2/27~\cite{Carolan711} and 1/4~\cite{PhysRevA.76.052326,PhysRevA.77.062316}.
Recent work~\cite{fldzhyan2021fivemode} proposed Bell states heralding with
5-port device and success rate 1/9 at the expense of photon-number resolving detection.
In this Section we confirm apparent optimality of $P=1/9$ circuit of~\cite{fldzhyan2021fivemode}.
Then we show that with 6 modes, 4 photons and single particle measurements success rate 2/27 is suboptimal,
improved circuit demonstrates $\sim 5\%$ better performance.
Although the advantage is almost negligible and hence is of little practical value,
it seems to best illustrate the potential power of our approach.

In our notations generation of Bell pairs with 5 mode and 4 photons implies
$\ket{\psi_{in}}=\ket{11110}$, 
$\ket{\psi_{out}} = (\ket{1010} + \ket{0101})/\sqrt{2}$,
with heralding measurement pattern being $\ket{\vec{m}} = \ket{2}$.
To inspect the optimality of scheme of Ref.~\cite{fldzhyan2021fivemode} 
we performed 400 independent optimizations of this setup
with tolerances $\epsilon_R = \epsilon_T^2 = 10^{-12}$.
The results could be summarized remarkably short.
All runs finished flawlessly in less then 45 iterations, always yielding
small optical residual, $R \le \epsilon_R$.
Identified solutions exhibit only two success probabilities,
either $P = 0$ or $P = 0.11111$,
the former being most attractive: 344 runs finished at this trivial value,
while 56 cases revealed appropriate $P=1/9$ success rate.
Corresponding scattering matrices turned out to be identical to that of Ref.~\cite{fldzhyan2021fivemode}).
Note in passing that this example provides lucid illustration of design space
enlargement idea alluded to Introduction.
Namely, if one adds additional vacuum mode to the above setup
numerical properties of the problem become much more favorable.
In particular, probability to get the same $P=1/9$ solution rises up to $\sim 95\%$.


\begin{figure}[t]
\centerline{\includegraphics[width=0.5\textwidth]{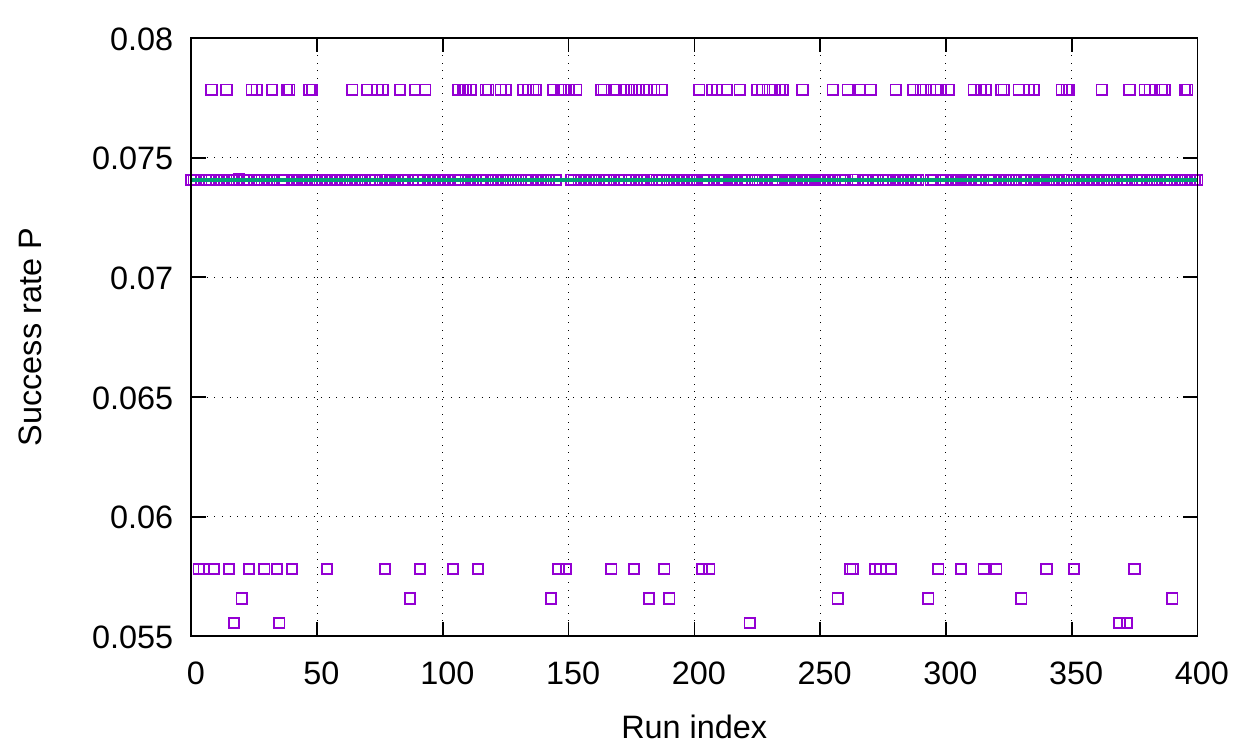}}
\caption{Maximal attained success probabilities of Bell pair generation with 4 photons and 6-port device
in 400 independent runs. Horizontal line represents conventional value $P = 2/27$.}
\label{fig::bell-6-4-all-runs}
\end{figure}

Next we examine the optimality of well-known 6 mode scheme~\cite{Carolan711}
with single photon detections.
Accordingly, input and output states to be considered are
\begin{align}
\label{bell-6-4-psi_out}
\ket{\psi_{in}} =\, & \ket{111100}\,,\nonumber \\
\ket{\psi_{out}} =\, & (\ket{0011} - \ket{1100})/\sqrt{2}\,,
\end{align}
while heralding measurement pattern is $\ket{\vec{m}} = \ket{11}$.
We conducted 400 independent runs within the stated setup
using $\epsilon_R = 10^{-10}$, $\epsilon_T = 10^{-6}$.
It turns out that in none of cases optimization finished in infeasible position,
all runs delivered appropriate solutions with non-zero success probabilities.
Fig.~\ref{fig::bell-6-4-all-runs} summarizes achieved success rates.
Apparently, the problem possesses five (locally) optimal points,
the one with $P = 2/27$ being most attractive.
However, the next frequent attraction level corresponds to $P \approx 0.07784$
and no doubt exceeds conventional value.
We ensured that excess cannot be explained by inexact numerics:
cited probability remains intact with tolerances reduced to
$\epsilon_R = \epsilon_T^2 = 10^{-14}$
and is insensitive to the parameters of involved algorithms.

We numerically extracted corresponding scattering matrices,
in most cases they are fairly described by the {\it ansatz}:
{\small
\beq
\label{bell-6m4ph-transform}
S = \left[
\begin{array}{rrrrrr}
\myrule  1/2    & -1/2   &  1/2    & -1/2    &  0      &  0      \\
\myrule  1/2    & -1/2   & -1/2    &  1/2    &  0      &  0      \\
\myrule  \alpha & \alpha & \beta   &  \beta  & -\gamma &  0      \\
\myrule  \beta  & \beta  & -\alpha & -\alpha &  0      & -\gamma \\
\myrule  \delta & \delta & -\sigma & -\sigma &  \nu    &  \mu    \\
\myrule  \sigma & \sigma & \delta  &  \delta &  \mu    & -\nu    \\
\end{array}
\right]
\eeq
}with all parameters being real positive. Note that unitarity requires specific normalization
of $\alpha, \beta, \delta, \sigma$ coefficients and uniquely determines $\mu, \nu, \gamma$ in terms of these.
Within the {\it ansatz}  \eq{bell-6m4ph-transform}
successful measurement of single photons in last two modes heralds the state
{\small
\begin{gather}
\left\{
(\delta^2 - \sigma^2)
[ 4 \alpha \beta (\createN{2}{2} - \createN{3}{2}) + 4 (\beta^2 - \alpha^2) \create{2}\create{3} ] +
\right. \\
\left.
2 \delta \sigma [ (\beta^2 - \alpha^2) ( \createN{2}{2} - \createN{3}{2} ) - 4 \alpha \beta \create{2}\create{3} + \create{0}\create{1}]
\right\}\,
\vac \nonumber
\end{gather}
}in ports 0-3.
With angle-like representation
$\alpha + i \beta = e^{i\phi} \cos\theta /\sqrt{2}$, $\delta + i \sigma =  e^{i\psi} \sin\theta/\sqrt{2}$
the absence of doubly occupied modes and similarity of
$\create{0}\create{1}$ and $\create{2}\create{3}$
amplitudes imply
$\mathrm{tg}(2\psi) = 2 \mathrm{tg}(2\phi)$, $\cos^2\theta = \sin(2\phi)$,
thus leaving single degree of freedom $\phi$.
In turn, the success probability becomes rational function of $x = \sin(2\phi)$
\beq
P(x) ~=~ 2 \, (1 - x)^2 \, x^2 \, / \, (1 + 3 x^2)
\eeq
the maximum of which occurs at
\begin{gather}
x^* = \frac{1}{3} ( \, \left[\frac{\sqrt{113} + 9}{2}\right]^{1/3} -
                       \left[\frac{\sqrt{113} - 9}{2}\right]^{1/3} \, ) \nonumber \\
\approx~ 0.40231994\,.
\label{bell-6-4-success-root}
\end{gather}

\begin{figure}[t]
\centerline{\includegraphics[width=0.5\textwidth]{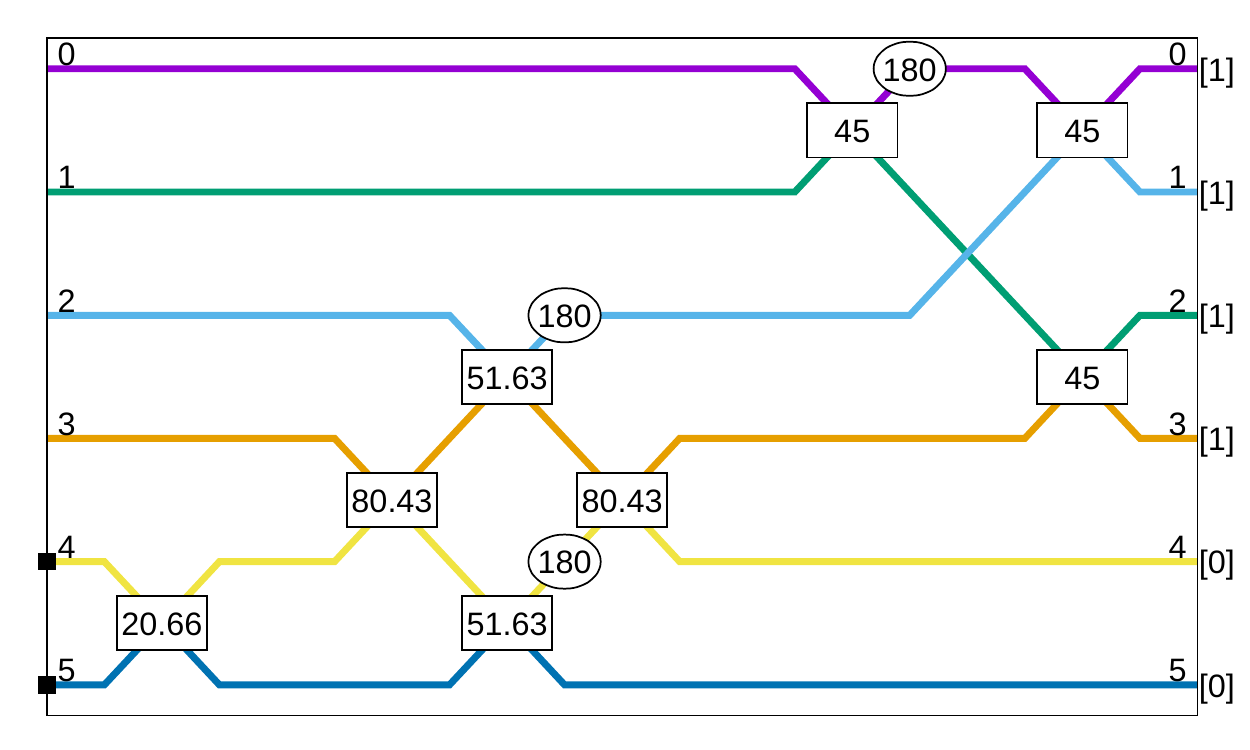}}
\caption{Decomposition of scattering matrix \eq{bell-6m4ph-transform} into elementary
optical components (see text for details).}
\label{fig::bell-6-4-split}
\end{figure}

\noindent Corresponding optimal success rate is approximately
\beq
\label{bell-6-4-success-rate}
P(x^*) ~\approx ~ 0.07784190
\eeq
in agreement with numerical estimates.
Therefore, analytic calculations confirm that success probability $2/27 \approx 0.074$ of 6-port device is suboptimal.
Larger rate is achievable although the advantage $\sim 5\%$ is almost negligible.
However, our point is not in practical improvement {\it per se}:
the result \eq{bell-6-4-success-rate} indicates that attempted
approach is able to correct even this widely-known well-studied case.

To conclude, it remains to consider the decomposition of \eq{bell-6m4ph-transform} into elementary optical
components (here we follow Ref.~\cite{Clements:16}).
After minor simplifications resulting scheme is presented in Fig.~\ref{fig::bell-6-4-split},
where each white square represents two-mode splitter
\beq
T(\theta) = \left[ \begin{array}{cc} \cos\theta & -\sin\theta \\ \sin\theta & \cos\theta \end{array}\right]
\eeq
with indicated approximate $\theta$-parameter in degrees, while ellipses stand for one-mode phase shifters
(phases are in degrees as well and are just the sign flips in the considered case).
Input photon counts are indicated in square brackets on the right,
while black square blobs represent heralding one-particle measurements.
Upon successful detection of single photons in output ports 4 and 5 (left hand side)
the scheme heralds Bell pair \eq{bell-6-4-psi_out} with success rate \eq{bell-6-4-success-rate}.

\section{Conclusions}

We developed an alternative approach to optimize linear optical heralding circuits.
Proposed method operates in entire Fock space and provides exact heralding of desired output.
Realizability by linear optics is assured by specific conditions put forward recently.
Respective optimization problem is well posed and bypasses difficulties inherent to conventional treatments.
We devised dedicated solution method, which exploits the structure of proposed formulation
and admits inexpensive implementation.
Prime drawback of considered methodology is limited applicability range:
Fock space dimensionality restricts the approach to moderately sized geometries,
which we fairly estimate to have no more than $\sim 10$ modes/photons.
As an application example we addressed the heralding of Bell states
and demonstrated that with 6 modes and 4 input photons
conventional $P=2/27$ scheme is suboptimal.
\bibliography{biblio}
\end{document}